# Newton's aether model

**Eric Baird**  (eric_baird@compuserve.com)

Isaac Newton is usually associated with the idea of absolute space and time, and with ballistic light-corpuscle arguments. However, Newton was also a proponent of wave/particle duality, and published a "new" variable-density aether model in which light and matter trajectories were either bent by gravitational fields, or deflected by an aether density gradient. Newton's (flawed) aether model can be considered as an early attempt at a curved-space model of gravity.

## 1. Introduction

Modern textbooks typically say that Newton believed that space and time were absolute and inviolable. However, a reading of Newton's "Principia" [1] and "Opticks" [2] reveals a rather different picture, with Optiks in particular documenting Newton's attempt to produce a model of gravity in which a gravitational field could be represented as a series of light-distance differentials, or as a variation in lightspeed or refractive index. This can be compared to Einstein's "refractive" approach to gravitational light-bending in 1911 ( [3] §4 ) and to his description of general relativity as a (nonparticulate!) gravitational aether model in 1920 [4][5].

We briefly look at some of the features of Newton's model, the mistake that doomed it to obscurity [6], and some of the consequences of this mistake on the subsequent development of physics.

## 2. Absolute space?

In "Principia", Newton was careful to distinguish between **relative space and time**, which were to be defined by observations and instrument readings, and **absolute space and time**, which were to relate to more abstract (and possibly arbitrary) quantities that might or might not have an identifiable grounding in physical reality. Finite-lightspeed effects had already been seen in the timing offsets in the orbits of Jupiter's moons, so this distinction was important. Newton insisted that the words "space" and "time" should by default refer to "absolute" (deduced, mathematical) quantities rather than their "apparent" counterparts, but statements from Newton regarding "absolute space" and "absolute time" do not automatically mean that Newton believed that directly-measurable distances and physical clock-rates were also absolute – (Principia, Definitions: "*… the natural days are truly unequal, though they are commonly considered as equal … it may well be that there is no such thing as an equable motion, whereby time may be accurately measured.*").

In "Opticks", Newton's idealised absolute space is occupied by a "new" form of medium whose density depends on gravitational properties, with variations in aether density producing the effects that would otherwise be described as the results of a gravitational field. The resulting metric associates a gravitational field with signal flight-time differences (*see:* Shapiro effect) that deflect light, leading to a normalised lightbeam-geometry that is not Euclidean. Since these effects are described in modern theory as the effects of curved space, it seems reasonable to interpret Newton's "absolute space" as an absolute Euclidean embedding-space that acts as a container for non-Euclidean geometry, rather than as an indication that Newton believed that gravity had no effect on measured or perceived distances, times, or "effective" geometrical relationships.

## 3. Lightspeed problems

Newton and Huyghens had opposing ideas on how a lightspeed differential deflected light:

**Newton view:**
"A gravitational gradient is associated with a change in speed of freely falling particles, with the speed being higher where the gravitational field is stronger. The bending of light at an air-glass boundary and the falling of light-corpuscles in a gravitational field can be described as the deflection of light towards regions of higher lightspeed."

**Huyghens view:**
"If a region has a slower speed of light, it will tend to collect light from the surrounding region. If a light-signal wavefront encounters a lightspeed gradient across its surface, with a faster natural speed on one side and a slower speed on the other, the retardation of the wavefront's "slower" side will steer the wavefront towards the slower-speed region."

Measurements of relative lightspeeds in different media in the 19th Century showed that it was Huyghens' explanation that was correct.

## 4. Huyghens' principle

Huyghens' principle is illustrated in the following diagrams of a light plane-wave hitting a glass block:.

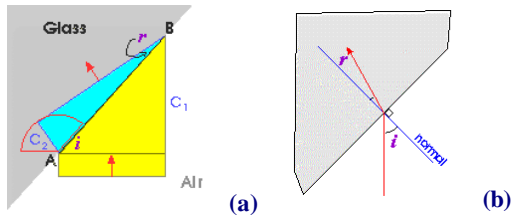

(a)         (b)

Figure (a) shows the progress of a wavefront hitting a lightspeed transition boundary, (b) shows the resulting change in direction of the wave normal.

At the start of the experiment, the leftmost edge of the advancing wavefront hits the air/glass boundary at **A**. A short time later, the rightmost edge of the wavefront has advanced by a distance $c_1$ and the rightmost edge of the wavefront has reached the boundary at **B**. By this time, the leftmost edge has penetrated the glass a smaller distance $c_2$ (because lightspeed in glass is slower) and the new wavefront lies along a line between **B** and a tangent centred on **A** with radius $c_2$. The wavefront normal is deflected to point more towards the region of slower lightspeed, with the exact relationship being

$$\frac{\sin_{INCIDENCEANGLE}}{\sin_{REFRACTIONANGLE}} = \frac{c_1}{c_2}$$

## 5. The "corrected" gravitational aether

We can make Newton's description compatible with Huyghens' principle by inverting his lightspeed and aether-density relationships. After these substitutions, the relevant queries in Opticks read:

**Qu 19x (rewritten)**

> Doth not the Refraction of Light proceed from the different density of this Aetherial Medium in different places, the Light receding always from the [rarer] parts of the Medium? And is not the density thereof [less] in free and open spaces void of Air and other grosser Bodies, than within the Pores of Water, Glass, Crystal, Gems, and other compact Bodies? …

**Qu. 21x (rewritten)**

> Is not this Medium much [denser] within the dense Bodies of the Sun, Stars, Planets and Comets, than in the empty Celestial spaces between them? And in passing from them to great distances, doth it not grow [rarer] and [rarer] perpetually and thereby cause the gravity of those great Bodies towards one another, and of their parts toward the Bodies; every Body endeavouring to go from the [rarer] parts of the Medium towards the [denser]? …
> … And though this [Decrease] of density may at great distances be exceeding slow, yet is the elastick force of this Medium be exceeding great, it may suffice to impel Bodies from the [rarer] parts of the medium towards the [denser], with all that power which we call Gravity. …

## 6. Some other issues

Opticks' **Query 1**, on light-bending effects:

> Do not Bodies act upon Light at a distance, and by their action bend its Rays; and is not this action strongest at the least distance?

This query does not make a distinction between gravitational light bending (action of gravity on unspecified "corpuscles", mentioned in Principia), and more conventional lensing effects. **Query 4** makes conventional optical effects more "gravitational" by proposing that "… rays of Light … reflected of refracted, begin to bend before they arrive at the Bodies …"

**Query 17**: Total internal reflection at a glass-air boundary introduced the philosophical problem of how the behaviour of light in glass could be affected by properties of a region that the light did not actually reach. How does light "know" what is beyond the glass, if it never actually passes beyond the glass? Newton's answer – that there must also be a hyper-fast wave-effect whose interference patterns then steer the subsequent light signal – predates the "pilot wave" description of the two-slit problem in quantum mechanics.

**Query 21** introduces the supposition that the aether might be particulate, but includes a slight qualification: "… (for I do not know what this *Aether* is) …".

**Query 28** recognises that lightwaves are not compression waves in the gravitational medium, since a compression-wave would have a tendency to spread out into less compressed regions (a lightbeam would then be deflected towards "dark" regions).

Newton's perplexity at how a single medium could then support both gravitational signals and electromagnetic signals can be compared to Einstein's similar musings in 1920 ("… two realities that are completely separated from each other conceptually, although connected causally, namely, gravitational ether and electromagnetic field …". [4]

**Quest. 30** (sic) asks whether light and matter are not interconvertible, and **Quest. 31** touches on the idea of stronger short-range forces being at the heart of chemical reactions.

## 7. Effective curvature

In the diagrams below, **(a)** shows genuinely flat space, **(b)** shows the aether-density gradient associated with the "corrected" version of Newton's variable-density aether model, and **(c)** shows the same map, extruded so that light-distances can be measured directly from map distances in the map:

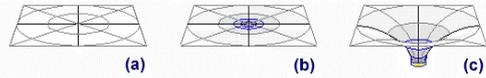

**(a) :truly flat space, (b): variable-density aether on flat background, (c): normalised light-distance map**

The physics of maps **(b)** ("aether-density gradient") and **(c)** ("curved space") can be equivalent (*see:* Thorne [7], Chapter 11 "What is Reality?").

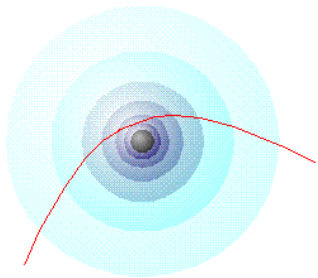

**Gravitational field approximated as a series of shells of increasing refractive index**

Although Cavendish did not calculate the sun's light-bending effect until the late 18$^{th}$ Century, Newton had already made similar calculations for the effects of a variable-density medium, in order to predict optical effects caused by the Earth's (variable density) atmosphere [6] ).

## 8. Effective closure

Our "corrected" version of Newton's aether model (with the time taken by light to cross a distance increasing or decreasing with gravitational field strength), allows a nominally-infinite universe containing a "central" concentration of matter to appear to its inhabitants as a closed hyperspherical universe with an even distribution of matter and no distinguishable centre [8] – figure 2 of Einstein's "Geometry and Experience" lecture [9] gives a method of mapping between these two equivalent descriptions.

## 9. Historical consequences

The advantage of expressing a gravitational field as a variation in density of an underlying medium was that wavefronts and particles would then be deflected by the same amount (w/p equivalence, gravity as a "spatial density" effect). John Michell's 1783 letter to Henry Cavendish was able to build on the arguments in Principia and Opticks and conclude that light climbing out of a gravitational well should lose energy, with the image of a high-gravity star viewed through a prism being offset towards the weaker end of the spectrum [10]. Michell's paper also calculated the R=2M event horizon radius, and discussed the "modern" method of finding non-radiating stars from the motions of their "normal" companions. Michell did not describe distance-dependent signal flight-time differences in his double-star scenario (such as those expected in simple ballistic-photon theory superimposed on flat space, discounted by deSitter in 1913 [11][12][13]), possibly because of uncertainty as to whether Newton's aether should support multiple superimposed lightspeeds. LaPlace also derived the R=2M relationship, and Cavendish and Soldner both calculated "Newtonian" values for the Sun's bending of light (*see:* [14][15][16] and Thorne [7] pp.122-123 & 132-133).

The disagreement between Newton's and Huyghens' arguments (Section 3), and the subsequent disproof of Newton's lightspeed predictions (Foucault, 1850) had serious consequences for the idea of wave-particle duality, with Optiks going out of print until 1931, and its contents apparently unknown to Einstein as late as 1921[4]. Other related work suffered a similar fate – laPlace removed his reference to the r=2M radius in later editions of his book (Thorne [7] p.122-123), Cavendish's calculation of solar light-deflection did not find its way into print until 1921, and Michell's paper dropped out of the citation chain and was only "rediscovered" in about 1979 [17][18]. After Laplace and Soldner's published "light-corpuscle" pieces in 1799 and 1801, there seems to be a "gap" in the reference chain until Einstein's 1911 paper. Even after Riemann's groundbreaking work on non-Euclidean geometry [19], attempts to construct curved-space models were not always taken seriously (e.g. Clerk Maxwell on W.K. Clifford's work, ~1869, "the work of a space-crumpler" [20]).

Spatial-curvature models remained problematical until after Einstein had repeated Michell's gravity-shift exercise (apparently oblivious to most or all of these earlier works!) and concluded that the situation could not be resolved unless an increased gravitational field was also associated with a reduction in the rate of timeflow. By arguing that gravity distorted maps of timeflow across a region, Einstein then opened the door to models of space*time* curvature based on Riemann's geometry, the most famous being his own general theory.

## 10. Conclusions

Newton's aether model arguably represents one of the most serious missed opportunities in the history of gravitational physics.

Newton's repeated attempts to unify various branches of physics led him to the concept of wave/particle duality and to a model of gravity in which the gravitational field could be described as a density gradient, and in which the deflection of light or matter by the field was modelled as the effect of a variation in refractive index. In singly-connected space, this approach can be topologically equivalent to a curved-space model of gravity [7] (by contrast, general relativity is a curved space*time* model of gravity).

However, Newton's model inverted a key lightspeed relationship. Instead of being a description in which the gravitational field *itself* was the medium (to misquote Marshall McLuhan, "The medium is the metric.", *see also:* Einstein: "the aether of general relativity" [4] "If we imagine the gravitational field … to be removed … no 'topological space'." [21]), Newton's model produced a description of a gravitational medium that was *displaced* by the gravitational field, and this led to the model and its associated principles and predictions (such as wave/particle duality, gravitational light-bending and gravitational shifts) being largely forgotten until Einstein's 20[th] Century work on quantum mechanics and general relativity.

Rindler has already pointed out that the mathematical machinery for general relativity was available in the Eighteenth Century [22]. Given that Michell's gravity-shift prediction was tantalisingly close to being a prediction of gravitational time dilation (the effect missing from 18[th] Century curved-space models), it seems that the loss of Newton's aether model may have significantly held back the progress of gravitational physics.